\documentclass[twocolumn,showpacs,preprintnumbers]{revtex4}
\usepackage{amsfonts}
\usepackage{amsmath}
\usepackage{graphicx}
\usepackage{dcolumn}
\usepackage{bm}

\input{tcilatex}

\begin{document}

\preprint{}
\title{High energy cosmic rays from AGN and the GZK cutoff}
\author{Yukio Tomozawa}
\affiliation{Michigan Center for Theoretical Physics, Randall Laboratory of Physics,
University of Michigan, Ann Arbor, MI. 48109-1040}
\date{\today }

\begin{abstract}
Based on a model for the emission of high energy cosmic rays from AGN
(Active Galactic Nuclei) that has been proposed by the author, he reviews
the status of the GZK cutoff and the correlation of high energy cosmic ray
sources with AGN locations in the existing data. The determination of mass
for the incident particles seems to be a key factor, and a suggestion for
doing that has been made in this article.
\end{abstract}

\maketitle

\section{Introduction}

High energy gamma rays from AGN have been reported by\ the Compton
Observatory \cite{compton} and by Cerenkov detectors\cite{cerenkov}, and
high energy cosmic rays have been correlated with AGN\cite{auger}. However,
the status of the GZK cutoff\cite{gzk} has been a mystery for the past
several decades\cite{mystery}. The absence of the cutoff in the Akeno-AGASA
detector\cite{agasa} and its presence in the Fly's Eye collaboration\cite%
{flyeye}, the Yakutsk\cite{yakutsk} and the Pierre Auger Project\cite{auger}
present a puzzle. In this article, using the machinery of a model proposed
by the author since 1985, a possible resolution for this puzzle is
suggested. A further suggestion for the determination of the mass parameter
for the incident particles is proposed as a crucial test.

\section{Summary of the model}

In a series of articles\cite{cr1}-\cite{cr9}, the author has presented a
model for the emission of high energy particles from AGN. The following is a
summary of the model.

1) Quantum effects on gravity yield repulsive forces at short distances\cite%
{cr1},\cite{cr3}.

2) The collapse of black holes results in explosive bounce back motion with
the emission of high energy particles.

3) Consideration of the Penrose diagram eliminates the horizon problem for
black holes\cite{cr4}. Black holes are not black anymore.

4) The knee energy for high energy cosmic rays can be understood as a split
between a radiation-dominated region and a matter dominated region, not
unlike that in the expansion of the universe. (See page 10 of the lecture
notes\cite{cr1}-\cite{cr3}.)

5) Neutrinos and gamma rays as well as cosmic rays should have the same
spectral index for each AGN. They should show a knee energy phenomenon, a
break in the energy spectral index, similar to that for the cosmic ray
energy spectrum.

6) The recent announcement by Hawking rescinding an earlier claim about the
information paradox\cite{hawking} is consistent with this model.

Further discussion of the knee energy in the model yields the existence of a
new mass scale in the knee energy range\cite{crnew}. The following are
additional features of the model.

7) If the proposed new particle with mass in the knee energy range (0.1 PeV$%
\sim $2 PeV) is stable and weakly interacting with ordinary particles, then
it becomes a candidate for dark matter. It does not necessarily have to be a
supersymmetric particle. That is an open question. However, if it is
supersymmetric, then it is easy to make a model for a weakly interacting
particle\cite{pevss}. The only requirement is that such particles must be
present in AGN or black holes so that the the knee energy is observed when
cosmic rays are emitted from AGN. A suggested name for the particle is
kneeon, sion (xion) or hizon\cite{crnew}.

8) If the particle is weakly interacting, then it does not obey the GZK
cutoff, since its interaction with photons in cosmic backgroud radiation is
weak. This is a possibe resolution of the GZK puzzle.

\section{ Resolution of the GZK puzzle}

We assume that the incident particles above the GZK cutoff observed by the
Akeno-AGASA detector are weakly interacting particles at the PeV mass scale,
which are required to exist in order to explain the phenomenon of the cosmic
ray knee energy in the model. One has to explain a mechanism whereby the
Akeno-AGASA detector is sensitive to such weakly interacting particles and
all other detectors are not. This is quite conceivable, since the spacing of
the detectors in the Akeno-AGASA apparatus is small (1 km between two
detectors) compared with that of the other detectors\cite{jones} (1.5 km
between two detectors for the Pierre Auger Project). Since weakly
interacting particles in high energy cosmic rays tend to make showers at a
lower altitude of the atmosphere due to the smaller cross sections, the
Akeno-AGASA detector is expected to observe higher percentage of weakly
interacting particles. Leaving this task of quantitative estimate to the
experimentalists, the author suggests that experimental groups to determine
the mass value of the incident particles at least enough to discriminate
whether they are ordinary particles (protons or nuclei) or have heavy mass
in the PeV range. In the following, the author presents a simple program to
determine the mass of the incident particles.

First let us consider the simple process $\pi ^{0}\rightarrow 2\gamma $.
Assume that the $2\gamma $ decay is perpendicular to the direction of the $%
\pi ^{0}$ in the rest frame of the pion and the energy and the angle of $%
2\gamma $ in the lab frame be $\omega $ and $2\theta $. Then, the total
energy and momentum of the system are $2\omega $ and $2\omega \cos \theta $
respectively, and the mass of the incident pion is given by%
\begin{equation}
mass=\sqrt{(2\omega )^{2}-(2\omega \cos \theta )^{2}}=2\omega \sin \theta .
\end{equation}%
In other words, the spread in the perpendicular momentum is a measure of the
mass of the incident particle. Assuming that all the secondary particles are
ordinary particles, i.e., massless particles compared with the PeV energy
scale, the rest mass of the incident particle is estimated by%
\begin{equation}
mass=\sqrt{(\dsum\limits_{j}E_{j})^{2}-(\sum_{j}E_{j}\cos \theta _{j})^{2}},
\end{equation}%
where $E_{j}$ and $\theta _{j}$ are the energy and angle relative to the
direction of the incident particle for each component, respectively. We are
assuming that each component is an ordinary particle and therefore massless
relative to the PeV mass scale. If there is a subsystem that is like a jet
so that it is difficult to separate into components, one has to treat this
subsystem as a single object. In such a case, one has to determine the
energy and direction of the subsystem.

The key issue is whether the Akeno-AGASA data for incident particles above
the GZK cutoff have the PeV mass scale or not. That may be an opportunity to
find a new particle. Intuitively, one could ask whether their data have
large energy spreads. However, the determination of mass in the PeV range
for the incident particle becomes more difficult for higher energy, almost
impossible in the GZK cutoff range. The author suggests the following
program as an alternative strategy.

1) Start with the analysis of air showers with the lowest possible energy
around the knee energy. Determine the mass of the incident particles by the
above method and determine the characteristics of such high mass events. A
study of the correlation of energy and spread angle for the secondaries,
characteristic of muon components etc., might be the important issues. This
might suggest a signature for high mass events corresponding to new
particles among the incident primaries.

2) Find the fraction of high mass events for higher energy. Combining that
with a characteristic signature for high mass events might suggest how to
determine the desired events. The most likely signature is the correlation
of energy and angle for the secondaries.

3) Extension of the low energy data for this high mass event fraction to
higher energy might suggest whether the proposed resolution is likely\ true
or not. I.e., if this fraction is a constant or increasing function of
energy, then it is very likely true. By continuity argument, one might be
able to find whether the events that create the GZK cutoff violation are
weakly interactiong high mass particles.

This analysis is suggested for any experimental group doing high energy
cosmic ray measurements. Such a program could lead to the discovery of a new
particle, possibly a dark matter particle. That might go as follows.

I) The mass parameter for incident particles of cosmic ray showers near the
knee energy is determined. \ A particle in the PeV mass range is expected.
This would be the discovery of a sion (xion) or kneeon, which is responsible
for the knee energy phenomenon in high energy cosmic rays emitted from AGN.
Data from detectors for low energy showers above the knee energy, Akeno\cite%
{akeno}, Tibet\cite{tibet}, DICE\cite{dice}, CACTI\cite{cacti}, HEGRA\cite%
{hegra}, HiRes/MIA\cite{hires}, KASCADE\cite{kascade} and Yakutsk\cite%
{yakutsk} would be useful for the analysis.

II) A relationship between the observed high mass events and other
signatures such as energy-angle correlation is established. Extention of
this relationship to higher energy could establish that events violating the
GZK cutoff are due to high mass particles established in I). This would be
evidence that such particles are indeed dark matter particles.

\section{Invisible AGN}

Another problem among the data of the existing experimental groups is
whether the correlation between high energy cosmic rays and AGN is
positively established or not. In a recent article, the author has suggested
the possiblity of black holes made of dark matter, which results in
invisible AGN\cite{crnew}. Such an object is not noticed as an ordinary AGN,
so that the correlation is not recognized. But, if one surveys gamma ray
emitters, one should consider correlations with them as well. At the present
time, the correlation has been recognized more in the data of the Pierre
Auger Project than in that of other groups. It is unfortunate that the
northern hemisphere component of the Pierre Auger Project has not be
completed. One has to wait for a few more years to see whether the disparity
between the correlations in the northern and southern hemisphere is real.
More importantly, the correlation between the gamma ray emitters and high
energy cosmic rays should be analyzed.

There is other observational data where a north-south asymmetry is observed%
\cite{center}. This is the distribution of circular galaxies. An observed
asymmetry is explained by the presence of a center for the expansion of the
universe. A consistent result was obtained from the analysis of the cmb
(cosmic microwave background) dipole\cite{center}. It is conceivable to have
a disparity in the distribution of dark matter due to the presence of the
center of the universe and as a result, a disparity in the distribution of
invisible AGN in both hemispheres.

\bigskip

\begin{acknowledgments}
\bigskip The author would like to thank Lawrence W. Jones and Jean Krisch
for useful discussion and David N. Williams for reading the manuscript.
\end{acknowledgments}

\end{document}